\documentclass{aa}  
\bibpunct{(}{)}{;}{a}{}{,} 
\usepackage{graphicx}
\usepackage{txfonts}
\begin{document} 

\title{Nonlinear Alfvén wave dynamics at a 2D magnetic null point: ponderomotive force}
   

   \author{J.~O. Thurgood
          \and
          J.~A. McLaughlin
          }

   \institute{Department of Mathematics \& Information Sciences, Northumbria University, Newcastle Upon Tyne, NE1 8ST, UK \\
              \email{jonathan.thurgood@northumbria.ac.uk}
}
   \date{Received February 21$^\textrm{st}$, 2013; accepted May 24$^\textrm{th}$, 2013}

 
  \abstract
   {In  the linear, $\beta=0$ MHD regime, the transient properties of MHD waves in the vicinity of 2D null points are well known. The waves are decoupled and accumulate at predictable parts of the magnetic topology: fast waves accumulate at the null point; whereas Alfv\'en  waves cannot cross the separatricies.
    However, in nonlinear MHD mode conversion  can occur at regions of inhomogeneous Alfv\'en speed,
 suggesting that the decoupled nature of waves may not extend to the nonlinear regime.
   	}
   { We investigate the behaviour of low-amplitude Alfv\'en waves about a 2D magnetic null point in nonlinear, $\beta=0$ MHD.
   }
   {We numerically simulate the introduction of low-amplitude Alfv\'en waves into the vicinity of a magnetic null point using the nonlinear LARE2D code. 
   }
   {Unlike in the linear regime, we find that the Alfv\'en wave sustains cospatial \emph{daughter disturbances}, manifest in the transverse and longitudinal fluid velocity, owing to the action of nonlinear magnetic pressure gradients (viz. the  \emph{ponderomotive force}). 
These disturbances are dependent on the Alfv\'en wave and do not interact with the medium to excite magnetoacoustic waves, although the transverse daughter becomes focused at the null point.
Additionally, an independently propagating fast magnetoacoustic wave is generated during the early stages, which transports some of the initial Alfv\'en wave energy towards the null point. Subsequently, despite undergoing  dispersion and phase-mixing due to gradients in the Alfv\'en-speed profile ($\nabla c_A \neq \mathbf{0}$) there is no further nonlinear generation of fast waves. 
   }
   {We find that Alfv\'en waves at 2D cold null points behave largely as in the linear regime, however they  sustain transverse and longitudinal disturbances - effects absent in the linear regime - due to nonlinear magnetic pressure gradients.}

   \maketitle
%

\section{Introduction}
Since the launch of solar satellites such as SDO, TRACE, Hinode, and STEREO, equipped with  sufficiently high-resolution and high-cadence instrumentation,  it has become established that MHD waves and oscillations are abundant throughout the coronal plasma
(see, e.g., Nakariakov \& Verwichte \citeyear{NK2005}; De Moortel \citeyear{ineke2005}; Banerjee et al. \citeyear{banerjee07}; Ruderman \& Erd\'elyi \citeyear{Ruderman2009}; Goosens et al. \citeyear{goosens2011}; McLaughlin et al. \citeyear{JamesGaryViktorRobertus}).
Consequently, it is clear that a well-developed theory of MHD waves is required to understand many ongoing coronal processes and dynamics. Due to the high degree of magnetic structuring in the atmospheric plasma, the medium in which these waves propagate is fundamentally inhomogeneous, leading to complex wave dynamics.

\emph{Magnetic null points}, which are locations where magnetic induction (hence Alfv\'en speed) is zero, occur naturally in the corona as a consequence of the distribution of isolated magnetic flux sources on the photospheric surface, and are predicted by magnetic field extrapolations such as Brown \& Priest (\citeyear{BrownPriest01}) and Beveridge et al. (\citeyear{Beveridge2002}). These null points are, like wave motions, prolific throughout the corona (Close et al. \citeyear{closeparnellpriest04}, Longscope \& Parnell \citeyear{loncopeparnell09}, and R\'egnier et al. \citeyear{regnierparnellhaynes08}, give rough estimates of $1.0-4.0 \times 10^4$ null points) and as such are a prime example of the extreme inhomogeneity that propagating MHD waves  encounter in the corona. Null points have been implicated at the heart of many dynamic processes, such as in CMEs (the \textit{magnetic breakout model}, e.g., Antichos \citeyear{antichos98}; Antichos et al. \citeyear{antichos99}) and in \textit{oscillatory reconnection} (e.g., McLaughlin et al. \citeyear{JamesOR2009}; McLaughlin et al. \citeyear{JamesOR2012}).
Study of MHD wave theory about null points therefore directly contributes to our understanding of wave propagation in realistic coronal plasmas.

The transient behaviour of \emph{linear waves} about null points, and its consequences for solar physics, has been extensively studied (see the review by McLaughlin et al. \citeyear{JAMESNULLREVIEW}).
A series of investigations into linear MHD wave propagation in the vicinity of 2D $\beta=0$ null points was carried out by McLaughlin \& Hood (\citeyear{MH2004}; \citeyear{MH2005};   \citeyear{MH2006A}). These studies give two key results for the linear regime: i) the fast and Alfv\'en waves accumulate at predictable regions of the null point topology, regardless of initial configuration; and 
ii) these wave modes remain distinct and decoupled, \emph{and do not interact}.

Fast magnetoacoustic waves are focused towards the null point due to refraction, resulting in the accumulation of current density and ohmic heating at the null point.
Linear 2D,  $\beta=0$ null points are  thus predicted as locations of preferential heating due to passing fast magnetoacoustic waves.
The Alfv\'en wave is found to accumulate along the separatricies, \emph{which it cannot cross.} 

Various studies that extend the 2D theory to $\beta\neq0$ and/or 3D (for example, McLaughlin \& Hood \citeyear{MH2006B}; Galsgaard et al. \citeyear{klaus03}; McLaughlin et al. \citeyear{james083dwkb}; Thurgood \& McLaughlin \citeyear{Me2012a}) 
all confirm that these two key features carry over. These extensions also add further dynamics; for example, considering $\beta\neq0$ introduces  the slow mode which interacts with the fast wave.

In nonlinear MHD, the nonlinear Lorentz force 
(sometimes referred to as the \lq{ponderomotive force}\rq{})
 is known to facilitate interaction between the MHD modes in certain inhomogeneous scenarios. 
A large body of work regarding the ponderomotive effects of waves in various MHD scenarios  has demonstrated that nonlinear Alfv\'en waves can generate magnetoacoustic waves as they propagate through regions of inhomogeneous Alfv\'en speed (such as, e.g., Nakariakov et al. \citeyear{NK97}, Nakariakov et al. \citeyear{NK98}, Verwichte et al. \citeyear{Erwin99}, Botha et al. \citeyear{gert2000}, Tsiklauri et al. \citeyear{tsiklauri2001}, McLaughlin et al. \citeyear{jamesphasemixing2011},  Thurgood \& McLaughlin \citeyear{Me2013_PMF}). The specific nature of ponderomotive mode conversion is dependent upon the gradients in the amplitude of the pulse 
and gradients in the Alfv\'en-speed profile. 
As such, in inhomogeneous magnetic topologies such as around  null points, ponderomotive effects have the potential to make a  significant impact upon the wave  dynamics, energy transport, and dissipation.

The behaviour of the nonlinear fast wave at a 2D magnetic null point was investigated by McLaughlin et al. (\citeyear{JamesOR2009}). The authors found that, for sufficiently large driving amplitudes,  magnetoacoustic shock waves develop, deform the null point and cause magnetic reconnection. The authors reported that in the nonlinear regime, some current can escape the null point, yet accumulation/heating still occurs (the shocks waves also heat the plasma).  This process  of \emph{oscillatory reconnection} has been subsequently studied by Threlfall et al. (\citeyear{Threl2012}) and McLaughlin et al. (\citeyear{JamesOR2012}). 

Galsgaard et al. (\citeyear{klaus03}) considered weakly nonlinear simulations of twisting motions about an azimuthally symmetric 2.5D null point and  observed a small amount of current accumulation at the null point (relative to larger current accumulation along the spine/fan, which is associated with the Alfv\'en wave in the linear regime). The authors suggest that this is due to nonlinear mode conversion from the Alfv\'en to fast magnetoacoustic mode; however, their study did not consider the transient dynamics of waves and their interaction, but rather the  current accumulation over time subject to an initial condition (i.e. they do not track the wave motions). Whilst the work of Thurgood \& McLaughlin (\citeyear{Me2013_PMF}) suggests that such nonlinear conversion could indeed be the explanation, it is unclear whether this is the case.

In this paper we address the question: \emph{how does the weakly nonlinear Alfv\'en wave behave in the vicinity of a 2D null point?} To do so, we numerically solve the cold-plasma MHD equations to simulate the nonlinear wave dynamics at a null point where a pure linear Alfv\'en wave is driven at the boundary, i.e. initially we ensure there is no fast wave present. The paper is structured as follows. In $\S \ref{equations}$ we describe the governing equations of the model, and in $\S \ref{coordsyst}$ we discuss the coordinate system used to distinguish between different MHD modes. In $\S \ref{numsol}$ we detail the numerical method and present the results of our simulations in $\S \ref{3.1}$. We discuss the nonlinear effects observed in our experiments in $\S \ref{NLE}$, and summarise in $\S \ref{conc}$.


\newpage

\section{Mathematical Model}\label{section:2}

\subsection{Governing equations}\label{equations}
We consider a plasma with dynamics described  by ideal, 2.5D $\beta=0$ MHD, with translational invariance in the $\hat{\mathbf{z}}$-direction; thus $\partial/\partial z =0$. The governing nonlinear MHD equations are
\begin{eqnarray}
\rho\left[\frac{\partial\mathbf{v}}{\partial t}+(\mathbf{v}\cdot\mathbf{\nabla})\mathbf{v}\right]&=& \left(\frac{\mathbf{\nabla}\times\mathbf{B}}{\mu} \right)\times\mathbf{B}\quad \nonumber \\
\frac{\partial\mathbf{B}}{\partial t}&=&\mathbf{\nabla}\times(\mathbf{v}\times\mathbf{B})\quad\nonumber\\
\frac{\partial\rho}{\partial t}&=& -\mathbf{\nabla}\cdot(\rho\mathbf{v})  \quad
\label{MHDeqns}
\end{eqnarray}
where the standard MHD notation applies: $\mathbf{v}$ is plasma velocity, 
$\rho$ is density, $\mathbf{B}$ is the magnetic field/induction,  $\gamma=5/3$ is the adiabatic index, and $\mu$ is the magnetic permeability. 
We consider an equilibrium state of $\rho=\rho_0$, 
(where $\rho_0$ is constant),
 ${\mathbf{v}}=\mathbf{0}$ and  equilibrium magnetic field ${\bf{B}}=\mathbf{B}_0$. Finite perturbations are considered in the form
$\rho=\rho_0 + \rho_1(\mathbf{r},t) $,
 $\mathbf{v}=\mathbf{0}+\mathbf{v}(\mathbf{r},t)$ and $\mathbf{B}=\mathbf{B}_0 +\mathbf{b}(\mathbf{r},t)$
and a subsequent nondimensionalisation using the substitution $\mathbf{v}=\overline{v}\,\mathbf{v}^*$,$\nabla={\nabla^*}/{L}$, $\mathbf{B}_{0}=B_{0}\mathbf{B}_{0}^{*}$, $\mathbf{b}=B_{0}\mathbf{b}^{*}$, $\mathbf{t}=\overline{t}\,\mathbf{t}^{*}$, $p_1=p_{0}p_{1}^{*}$ and $\rho_1=\rho_{0}\rho_{1}^{*}$ is performed, with the additional choices $\overline{v}={L} / {\overline{t}}$ and $\overline{v}={B_{0}} / {\sqrt{\mu\rho_{0}}}$.  The resulting nondimensionalised, governing equations of the perturbed system are
\begin{eqnarray}
\frac{\partial \mathbf{v}}{\partial t} &=& \left(\nabla\times\mathbf{b}\right)\times\mathbf{B}_0 + \mathbf{N}_{1} \nonumber\\
\frac{\partial\mathbf{b}}{\partial t} &=& \nabla\times\left( \mathbf{v} \times\mathbf{B}_{0}\right) +  \mathbf{N}_{2} \nonumber\\
\frac{\partial \rho_1}{\partial t} &=& -\nabla\cdot \mathbf{v} + {N}_{3}\nonumber\\
{\bf{N}}_1 &=& \left(\nabla\times\mathbf{b}\right)\times\mathbf{b} -\rho_{1}\frac{\partial \mathbf{v}}{\partial t}  - \left(1+\rho_1\right)\left(\mathbf{v}\cdot\nabla\right)\mathbf{v} \nonumber \\
{\bf{N}}_{2} &=& \nabla \times \left(\mathbf{v}\times\mathbf{b}\right)\nonumber \\
{{N}}_3 &=& -\nabla \cdot \left(\rho_{1} \mathbf{v} \right) 
\label{equation_MHD}
\end{eqnarray}
where terms ${\bf{N}}_i$ are the nonlinear components. The star indices have been dropped, henceforth all equations are presented in a nondimensional form.
The  equations are merged into one governing PDE
\begin{eqnarray}
\frac{\partial^2 \mathbf{v}}{\partial t^2}&=&\left\lbrace\mathbf{\nabla}\times\left[\mathbf{\nabla}\times\left(\mathbf{v}\times\mathbf{B}_0\right)\right]\right\rbrace\times\mathbf{B}_0 + \mathbf{N} \nonumber \\
\mathbf{N}&=&\left\lbrace\mathbf{\nabla}\times\left[\mathbf{\nabla}\times\left(\mathbf{v}\times\mathbf{b}\right)\right]\right\rbrace\times \left(\mathbf{B}_0+\mathbf{b}\right)
\nonumber \\
& & 
+\left(\nabla\times\mathbf{b}\right)\times    \left[\nabla\times\left(\mathbf{v}\times\mathbf{B}_{0}\right)\right] \nonumber \\
& & + \left(\mathbf{\nabla}\cdot\mathbf{v} - \mathbf{v} \cdot  \mathbf{\nabla}\right)\left(\mathbf{\nabla}\times\mathbf{b}\right)\times\mathbf{B}_0 \nonumber\\
& & - \rho_1 \left\lbrace\mathbf{\nabla}\times\left[\mathbf{\nabla}\times\left(\mathbf{v}\times\mathbf{B}_0\right)\right]\right\rbrace\times\mathbf{B}_0 \nonumber \\
& & -  \left[ \left(\mathbf{\nabla}\times\mathbf{b}\right)\times\mathbf{B}_0  \cdot\mathbf{\nabla}\right] \mathbf{v} \quad .
  \label{eqn:governingPDE}
\end{eqnarray}
The first term describes the linear regime of the system and the terms  $\mathbf{N}$ are the  nonlinear terms, displayed here to the second order for brevity (N.B. our solution solves the full system of equations, with all nonlinear terms, see $\S \ref{numsol}$ and Arber et al. \citeyear{LAREPAPER}).

\subsection{Isolating MHD modes}\label{coordsyst}

Thurgood \& McLaughlin (\citeyear{Me2012a}) developed a magnetic-flux-based coordinate system that allows the decomposition of MHD waves into constituent modes and the construction of initial conditions that correspond to single linear modes of oscillation.  This approach is suitable  for any  MHD scenario that is capable of sustaining true Alfv\'en waves - i.e. where the equilibrium configuration permits some invariant direction (see their section 2.3.1).

The projected perturbations corresponding to the wave modes according to this coordinate system are
\begin{eqnarray*}
\textrm{Alfv\'en wave }\mathbf{v}\mathrm{-perturbation} &:& v_z  \\
\textrm{Fast wave }\mathbf{v}\mathrm{-perturbation} &:&  {v_\perp} =
\mathbf{v}\cdot\hat{\mathbf{z}}\times\mathbf{B}_0 = -B_{y}v_{x}+B_{x}v_{y}\\
\textrm{Longitudinal }\mathbf{v}\mathrm{-perturbation} &:& {v_\parallel} =
\mathbf{v}\cdot\mathbf{B}_0 = B_{x}v_{x}+B_{y}v_{y} \quad .
\end{eqnarray*}
Perturbations 
 in the invariant direction elicit magnetic tension only and thus correspond to the Alfv\'en wave. Perturbations in the  direction transverse to both the equilibrium field and the invariant direction  (here the $xy$-plane) correspond to the fast wave, and perturbations in the longitudinal direction are static as there is no longitudinal force to transport the disturbance in our $\beta=0$ scenario (and would correspond to the slow wave in $\beta\neq0$).

Note that this corresponds to the coordinate system used in McLaughlin \& Hood (\citeyear{MH2004}) which is a specific implementation of the flux-based coordinate system appropriate for 2D magnetic null points.

\begin{figure}
\resizebox{\hsize}{!}{\includegraphics{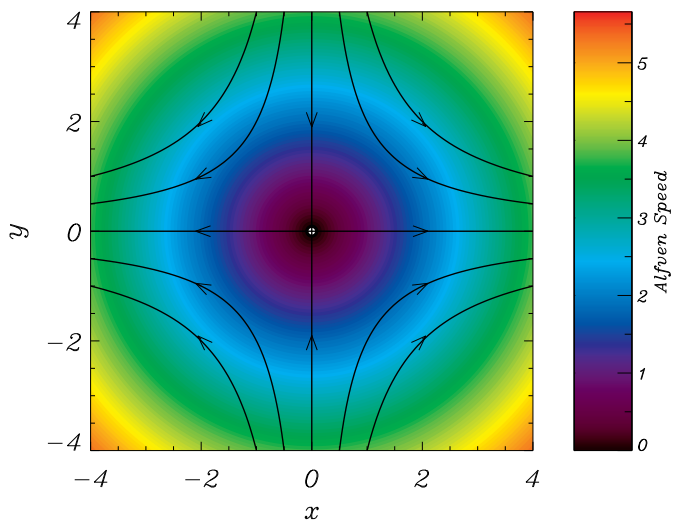}}
\caption{Indicative lines of the equilibrium magnetic field structure $\mathbf{B}_0$ (black lines) and contours of the Alfv\'en-speed profile $c_{A}=\sqrt{x^2+y^2}$ ($\rho_0$ is constant).}
\label{fig:fieldlines}
\end{figure}

\begin{figure*}
\sidecaption
\includegraphics[width=12cm]{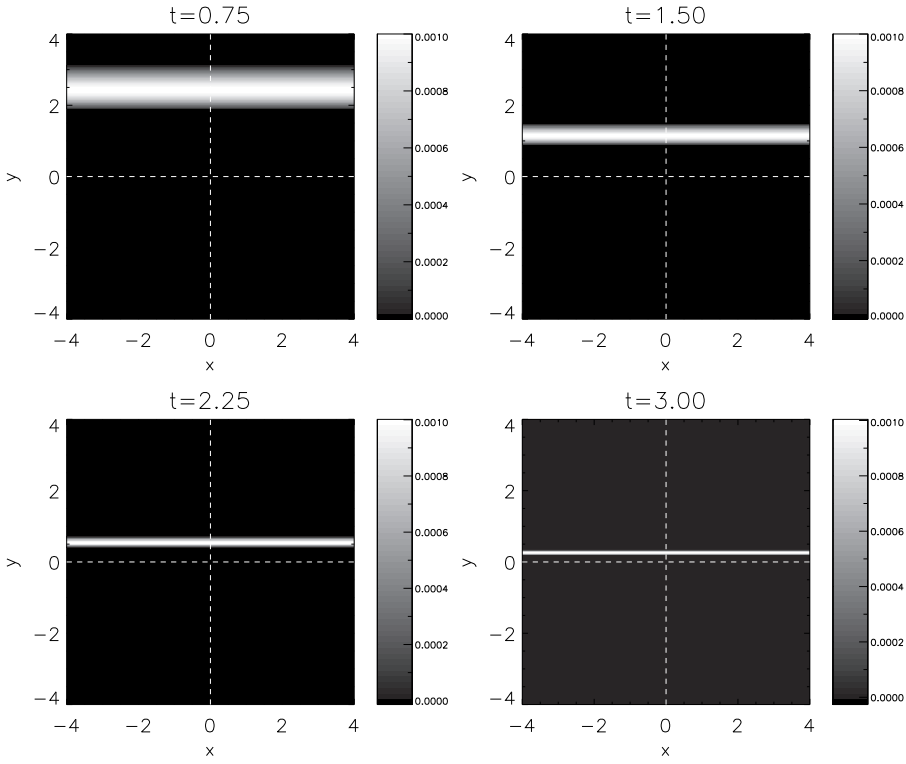}
\caption{The evolution of  $v_z$ (the Alfv\'en wave) over time. The Alfv\'en wave propagates along fieldlines at the Alfv\'en speed, hence we see a narrowing planar pulse propagating towards the $y=0$ separatrix (the separatricies are marked by dashed white lines).}
\label{weakvz}
\end{figure*}
\begin{figure*}
\sidecaption
\includegraphics[width=12cm]{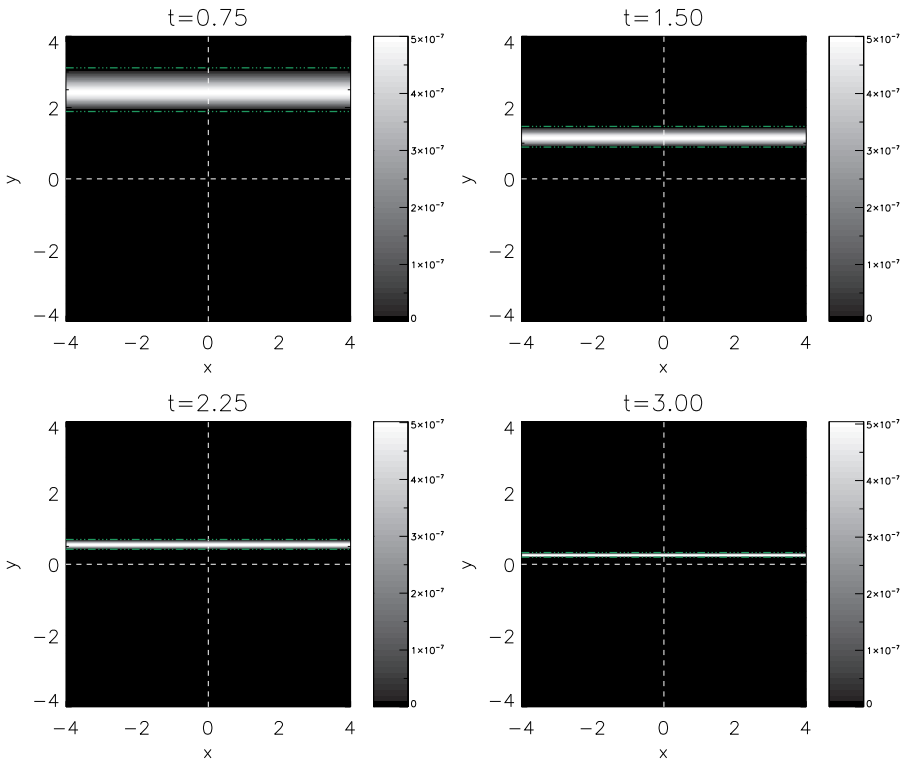}
\caption{The longitudinal component  $v_\parallel$ over time. We find that the ponderomotive force of the propagating Alfv\'en wave (position marked by green lines) sustains a cospatial disturbance in velocity along the background magnetic field. This \textit{longitudinal daughter disturbance} is a ponderomotive effect which in this case does not facilitate any conversion to the slow mode.}
\label{weakvpar}
\end{figure*}

\begin{figure*}
\centering
\includegraphics[width=17cm]{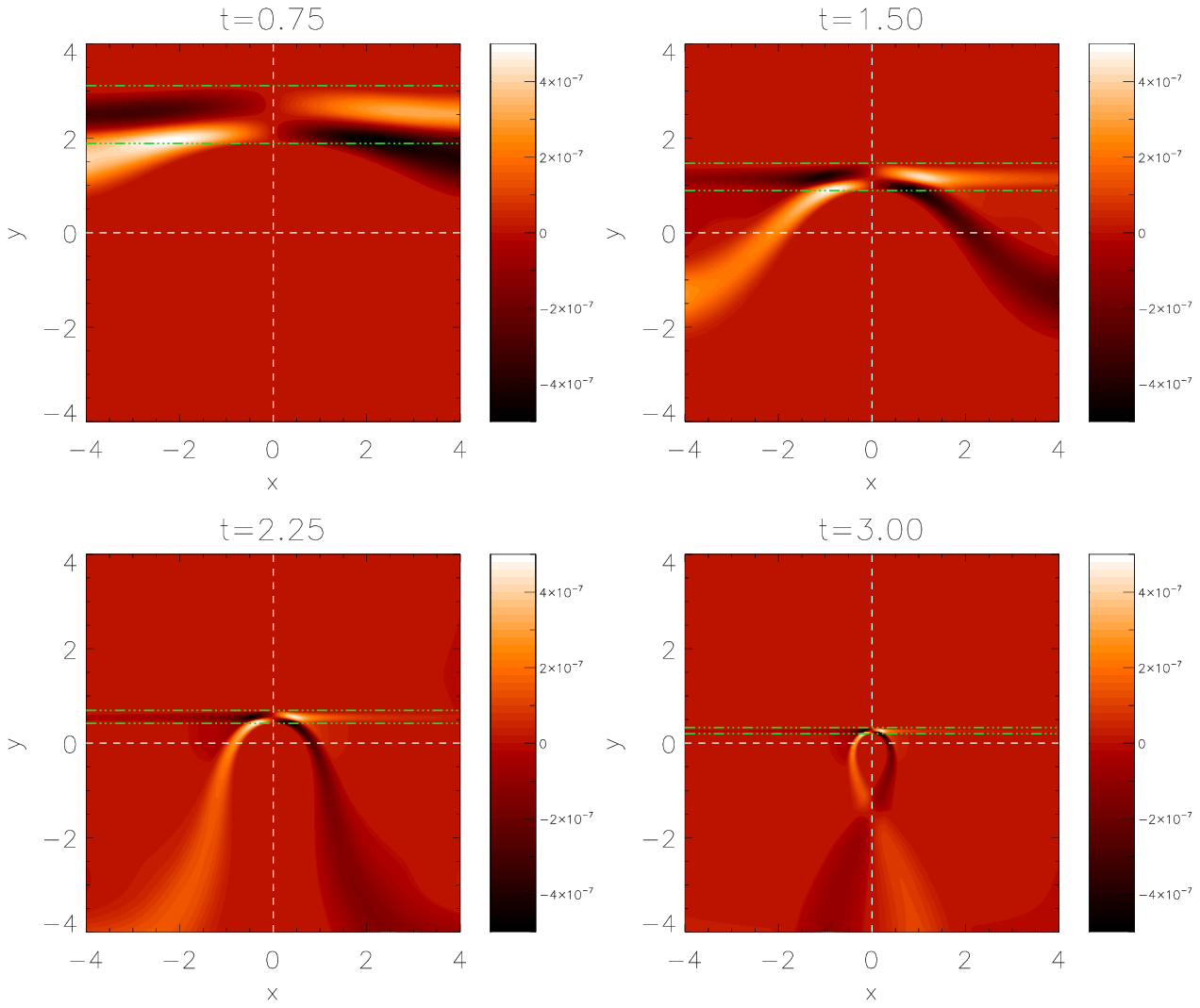}
\caption{The evolution of  $v_\perp$ over time. We find two pronounced nonlinear effects: a transverse daughter disturbance which is cospatial to the Alfv\'en wave (indicated by dashed green lines) and an independently propagating fast wave.}
\label{weakvperp}
\end{figure*}

\section{Numerical Simulation}\label{numsol}
We solve the full set of the nonlinear, nondimensionalised MHD equations (\ref{equation_MHD}) with the (nondimensionalised) equilibrium magnetic field 
\begin{equation}
\mathbf{B}_{0}=\left[x,-y, 0\right]  \label{eqn:2D_null_point}
\end{equation}
using the fully nonlinear, shock-capturing LARE2D code (Arber et al. \citeyear{LAREPAPER}).
Equation (\ref{eqn:2D_null_point}) corresponds to a 2D null point with two key topological features, the null point itself where magnetic induction is zero at the origin, and the separatrices at $x=0$ and $y=0$ lines (see Figure \ref{fig:fieldlines} and McLaughlin et al. \citeyear{JAMESNULLREVIEW}).
 We drive planar, sinusoidal pulses in ${v}_z$ and ${b}_z$ at the upper $y$-boundary to introduce a linear Alfv\'en wave as per $\S \ref{coordsyst}$. To do so we drive
\begin{equation}
v_{z}(x,4)=A\sin(2\pi t), \quad v_{x}=v_{y}=v_\perp=v_{\parallel}=0, \quad \mathbf{b}=-\sqrt{\mu\rho_{0}}\,\mathbf{v}
\label{drive_alfven}
\end{equation}
for $0\le t\le0.5$, and  we consider a driving amplitude $A=0.001$. This amplitude is small with respect to the characteristic velocity of the non-dimensionalisation used in $\S \ref{equations}$ ($\overline{v}={L} / {\overline{t}}$, i.e. a typical Alfv\'en crossing time over the length scale of interest) and thus we consider the weakly nonlinear scenario. Simple zero-gradient conditions are employed on the other boundaries, and the simulations are performed over the domain $x\in[-4,4]$, $y\in[-4,4]$ with $2400\times2400$ grid points.

\subsection{Results}\label{3.1}
In Figure \ref{weakvz}, we plot the propagating Alfv\'en wave in the velocity component $v_z$. We have computed the $b_z$ perturbation which shows a qualitatively identical result.
It is known that the linear Alfv\'en wave propagates along the magnetic field lines at the Alfv\'en speed $c_A$ (here $c_{A}=\sqrt{x^2+y^2}$), causing spreading of pulses along field lines and is unable to cross the separatricies  (McLaughlin \& Hood \citeyear{MH2004}). Given that our driving imposes a planar profile, this spreading effect is not obvious in our simulation, instead we see the planar pulse that propagates towards the $y=0$ separatrix at a speed which is equivalent to the Alfv\'en speed evaluated at $x=0$. It cannot cross the separatrix, as $c_{A}|_{x=0}\rightarrow 0$ as $y\rightarrow 0$, and it accumulates nearby with ever increasing gradients, hence resistive dissipation will eventually become an important consideration (McLaughlin \& Hood \citeyear{MH2004}). Thus,  \textit{$v_z$ behaves as in the linear regime.}

We now consider the other orthogonal velocity components ($v_\perp$ and $v_\parallel$) which remain zero throughout linear simulations, to investigate possible nonlinear effects.
We first consider the longitudinal velocity component $v_\parallel$, shown in  Figure \ref{weakvpar}. Here, we find a nonlinear disturbance of $\mathcal{O}(0.5A^2=5\times10^{-7})$, which is cospatial to the Alfv\'en wave pulse in time (the front and rear position of the Alfv\'en pulse is marked by green lines).  The pulse appears similar in profile to that of the Alfv\'en wave, however is more compressed and steep, i.e. the profile is approximately as a squared sine wave, whereas $v_z$ is sinusoidal. This disturbance is not an independently propagating wave (in $\beta=0$ such motion is prohibited), but a direct consequence of the longitudinal component of the ponderomotive force, which is induced, sustained and carried by the propagating Alfv\'en wave. This is the specific manifestation of the \emph{longitudinal daughter disturbance}, a general feature of nonlinear Alfv\'en waves and a common manifestation of the ponderomotive force (see Thurgood \& McLaughlin \citeyear{Me2013_PMF} for a detailed discussion). 

Now we consider the fast-mode velocity component $v_\perp$, shown in Figure \ref{weakvperp}, which consists of two features: a wave that propagates independently of the Alfv\'en wave and a cospatial disturbance, both nonlinear of order $\mathcal{O}(0.5A^2=5\times10^{-7})$. The independently propagating wave is generated during the driving stage of the simulation, and propagates with the transient characteristics of a linear fast wave at a single null point, namely that it undergoes refraction due to the Alfv\'en-speed profile, crosses the separatricies and accumulates at the null point. Hence, driving a linear Alfv\'en wave according to equation (\ref{drive_alfven}) has nonlinearly excited a fast wave.

We also find a disturbance in $v_\perp$ which does not appear to correspond to a fast wave (qualitatively, in terms of transient behaviour), and is cospatial to the Alfv\'en wave pulse (again, the position of which is shown by the green envelope in Figure \ref{weakvperp}). This is the \emph{transverse daughter disturbance}, detailed for general MHD in Thurgood \& McLaughlin (\citeyear{Me2013_PMF}). This disturbance is not an independently propagating wave. Within the cospatial region, the disturbance becomes increasing focused towards the separatrix and  ultimately to the vicinity of the null point.

\section{Nonlinear Effects}\label{NLE}
Here, due to the choice of small driving amplitude $A$, $v_z$ behaves as in the linear regime (Figure \ref{weakvz}). Whilst the choice of a low driving amplitude makes the nonlinear effects small, it is nonetheless sufficient to demonstrate in what ways the (shock-free) nonlinear system differs from the linear, in particular its interaction with the transverse and longitudinal fluid variables and between differing modes of oscillation. Here we see two types of nonlinear effects which are absent in the linear study of McLaughlin \& Hood ({\citeyear{MH2004}}), both of which are generated or sustained at $\mathcal{O}(A^{2}/2)$:daughter disturbances and independently propagating fast waves.

\subsection{Daughter disturbances}\label{conc_daughters}
In the simulations, we observe disturbances in $v_\perp$ and $v_\parallel$ which develop immediately and remain cospatial to the wave observed in $v_z$ throughout the simulations. The Alfv\'en wave exerts a nonlinear magnetic pressure gradient (viz. \emph{ponderomotive force}) upon the medium, resulting in these  \emph{transverse} and \emph{longitudinal daughter disturbances}. Thurgood \& McLaughlin (\citeyear{Me2013_PMF}) discuss the MHD-ponderomotive effects of Alfv\'en waves in detail, and show that such daughters will be sustained anywhere there are non-zero gradients in the pulse amplitude relative to the equilibrium magnetic field. At any given instant,  the ponderomotive force of an Alfv\'en wave manifest in $\hat{\mathbf{z}}$ across and along the field is (Thurgood \& McLaughlin \citeyear{Me2013_PMF}, section 3, equations 12-13)
\begin{equation}
\frac{\partial v_\perp}{\partial t}
=-\frac{1}{\mu\rho_0}\nabla_{\perp}\left(\frac{b_{z}^2}{2}\right) 
\longrightarrow \mathcal{O}\left(v_\perp\right) \sim \mathcal{O}\left(\frac{A^2}{2}\right)
\label{eq:PMF_perp}
\end{equation}
\begin{equation}
\frac{\partial v_\parallel}{\partial t}
=-\frac{1}{\mu\rho_0}\nabla_{\parallel}\left(\frac{b_{z}^2}{2}\right)
\longrightarrow \mathcal{O}\left(v_\parallel\right) \sim \mathcal{O}\left(\frac{A^2}{2}\right)
\quad .
\label{eq:PMF_par}
\end{equation}
Here, $\nabla_{\perp}\equiv\hat{\mathbf{z}}\,\times\mathbf{B}_{0} \cdot \nabla$ and $\nabla_{\parallel}\equiv \mathbf{B}_0 \cdot \nabla$ (these terms are the gradients transverse and longitudinal relative to the equilibrium magnetic field) and $b_{z}=\pm\sqrt{\mu\rho_{0}}v_{z}$. Where the net ponderomotive force over a wave period is non-zero, excitation of magnetoacoustic modes occurs. Where the action of the force of the leading pulse edge is consistently nullified by the trailing edge, disturbances arise in the transverse and longitudinal fluid-variables that do not excite wave motions but remain confined to a region cospatial to the Alfv\'en wave, referred to as the \emph{ponderomotive envelope} (see discussion of, e.g., \lq\lq{ponderomotive wings}\rq\rq{} in Verwichte et al. \citeyear{Erwin99}, and \lq\lq{daughter disturbances}\rq\rq{} in Thurgood \& McLaughlin \citeyear{Me2013_PMF}). 

In $\S \ref{numsol}$, the spatial distribution of the amplitude  of these two disturbances differs; the longitudinal daughter ($v_\parallel$) varies uniformly in $\hat{\mathbf{y}}$ with a similar profile to that of the Alfv\'en wave in $v_z$ (as the disturbance is generated at amplitude $A^2$, the profile is more akin to that of the squared sine wave, i.e. more compressed and steep). However, the transverse daughter ($v_\perp$) varies in both $\hat{\mathbf{x}}$ and $\hat{\mathbf{y}}$, is of opposite sign either side of the $x=0$ separatrix, and appears to become increasingly focused towards the $x=0$ separatrix as time evolves. 
By considering ($\ref{eq:PMF_perp}$) and ($\ref{eq:PMF_par}$) with the equilibrium field ($\ref{eqn:2D_null_point}$) and given that in the simulation the pulse remains planar in $\hat{\mathbf{x}}$ (i.e. has $\partial/\partial x =0 $ throughout), the transverse and longitudinal nonlinear magnetic pressure gradients (ponderomotive force) assume profiles such that
\begin{equation}
F_\perp
\sim
x\,\frac{\partial}{\partial y} \,\left(\frac{b_{z}^{2}}{2}\right)
\quad , \quad
F_\parallel
\sim
y\,\frac{\partial}{\partial y} \, \left(\frac{b_{z}^{2}}{2}\right)
\label{FperpFpar}
\end{equation}
where we know, from the numerical results where the leading edge propagates slower than the trailing (hence length scales decrease and gradients grow), that the derivative value becomes larger in time 
and the pulse profile changes in $y$, but that in $x$ does not. Transversely, within the ponderomotive envelope, there is an applied pressure gradient which  overall becomes stronger as the Alfv\'en wave tends towards the $y=0$ separatrix (as the gradient increases) yet maintains the same profile proportional to $|x|$ throughout, with zero magnetic pressure  along $x=0$. In the simulation, we see that the transverse daughter becomes increasingly focused towards $x=0$. Hence,  a possible explanation is that the applied magnetic pressure gradient acts to accelerate the fluid within the envelope towards $x=0$.
Longitudinally, as the leading edge of the Alfv\'en wave undergoes steepening the derivative value will be larger at the lead than that towards the trailing edge, however the value of $y$ is smaller. Since in the simulations we see no static perturbations in $v_\parallel$ or $b_\parallel$, the net force must be zero and thus the change in $y$ must be proportional to the increasing steepness of the leading edge throughout the simulation (this is intuitive, as net speed of the pulse in the $\hat{\mathbf{y}}$-direction also decreases proportional to $y$). Hence, we are confident that the observed phenomena are ponderomotive daughter disturbances, as they are generated at a nonlinear order and are consistent with equation (\ref{FperpFpar}). 

\subsection{Independently propagating fast wave}\label{conc_fast}
In $\S \ref{numsol}$ we have seen that a wave in $v_\perp$ is generated during the driving phase ($0\le t<0.5$).
This wave subsequently propagates with all of the transient features of a linear fast magnetoacoustic wave (refracting about the Alfv\'en speed profile and accumulating at the null point). As the nonlinear magnetic pressure is the only facilitator of coupling between $v_z$ and $v_\perp$ in cold, 2.5D MHD systems, the generation of such a wave is due to the exertion of a transverse ponderomotive force (equation \ref{eq:PMF_perp}) such that the net force over the period is non-zero. After the driving phase, no further excitation of fast waves occurs, thus the net ponderomotive force must be zero (and thus is only manifest in the aforementioned daughters). Hence, after this driving period gradients in the trailing edge nullify the fluid acceleration caused by the gradients in the leading edge.

The value of the net ponderomotive force can only change if the pulse geometry is altered. This requires a change in the Alfv\'en speed.
Here, gradients in the Alfv\'en speed are non-zero 
$$\nabla_{\perp} c_{A}=\frac{2xy}{\sqrt{x^2+y^2}}
\qquad , \qquad
 \nabla_{\parallel} c_{A}=\frac{x^2-y^2}{\sqrt{x^2+y^2}}$$
generally permitting such changes in the pulse geometry (i.e. our system is inhomogeneous). However since we impose a planar profile, the effective speed of the wave is $c_{A}|_{x=0}=y$, and  its derivative in the direction of propagation is constant. Thus, as the Alfv\'en wave propagates its geometry is not altered such that there should be a change from a non-zero to zero net ponderomotive force.

This raises the question that, since the net force \emph{cannot change}, why do we observe the generation of fast waves only during the driving phase, rather than continuously? There are two possible explanations

(i) {Physical Effect:} The net force is initially non-zero and remains so, i.e. there is a physical mechanism which suppresses the further generation of independently propagating fast waves.

(ii) {Mathematical Artefact:} The net force is actually zero, thus no ponderomotive excitation of the fast mode should occur. The excitation of the fast wave is a consequence of linearly driving a nonlinear system.

\subsubsection{Physical Effect}\label{PhysEff}
If the net-force of the wave is always non-zero, in the absence of continuous fast wave generation a mechanism must act to oppose further wave excitation.
Botha et al. (\citeyear{gert2000}) considered the case of a harmonic Alfv\'en wave propagating in a homogeneous field stratified by a transverse density profile. They reported that the transverse gradients caused the nonlinear excitation of fast waves. However, these waves  eventually saturated and did not continue to develop in time. This saturation was also later reported to occur for pulse-type Alfv\'en waves in the same equilibrium set-up by Tsiklauri et al. (\citeyear{tsiklauri2001}). Botha et al. (\citeyear{gert2000}) proposed that the saturation  occurs due  to wave interference between the generated fast waves. It is possible that in our system a level of saturation sufficient to oppose further mode conversion is reached so rapidly that only a single independent fast wave is generated.

\subsubsection{Mathematical Artefact}\label{MathA}
If the net ponderomotive force is zero throughout, then the independent fast wave has been introduced as a mathematical consequence of our driving condition. We have used the driving condition (\ref{drive_alfven}) so we can directly compare our nonlinear experiment to the linear results of McLaughlin \& Hood (\citeyear{MH2004}).  However, in nonlinear MHD, driving the $\hat{\mathbf{z}}$-components of the fluid variables and holding the transverse and longitudinal components at zero  corresponds to an Alfv\'en wave  with no instantaneous ponderomotive force.
  The driven Alfv\'en wave  subsequently enters an inhomogeneous region 
     which alters the pulse profile and contributes to the nonlinear magnetic pressure perturbation, unopposed by other factors (which are specified as zero). Thus 
  a non-zero ponderomotive force acts
   \emph{across} the field, resulting in the fast wave, and \emph{along} the field, resulting in a static longitudinal perturbation, as $\beta=0$ (this is subsequently removed by our boundary post-driving boundary conditions, hence absent from our discussion in $\S \ref{numsol}$). 

If this is indeed the case, then physically boundary condition (\ref{drive_alfven}) is inappropriate as it corresponds to an incoming wave with no ponderomotive daughters, and hence no ponderomotive force. At the bare minimum, longitudinal daughters are present for Alfv\'en wave pulses in straight-field, homogeneous MHD (see Verwichte et al. \citeyear{Erwin99}). From the perspective of wave-stability the driving conditions used are entirely appropriate in the linear regime, as they correspond to a pure linear Alfv\'en wave (i.e. a wave driven solely by magnetic tension, as per Alfv\'en \citeyear{Alfven42})  that does not interact with other modes of oscillation (see Parker \citeyear{Parker91}). This can be be confirmed by considering equation ($\ref{eqn:governingPDE}$) with $\mathbf{N}=\mathbf{0}$ in the flux-based coordinate system, which yields three separate and decoupled equations for invariant, transverse and longitudinal variables, i.e. for the Alfv\'en, fast and (absent) slow modes of oscillation. As perturbations in $\hat{\mathbf{z}}$ do not elicit responses in other directions in the linear regime, such a wave could be considered \emph{linearly stable.} 
However, in the full nonlinear system, disturbances to $v_z$ do not exist independently to perturbations in $v_\perp$ and $v_\parallel$ due to the action of the ponderomotive force. The driving conditions still successfully introduce an Alfv\'en wave (in the nonlinear regime the motion of an Alfv\'en wave is still primarily due to linear magnetic tension), yet specify values of transverse and longitudinal fluid variables that are inconsistent with those specified by the equations for a single Alfv\'en wave (these values should correspond to those implied by  equations \ref{eq:PMF_perp} and \ref{eq:PMF_par}).

In the absence of a physical saturation mechanism such as that described in $\S \ref{PhysEff}$, the independent fast wave is a consequence of linearly driving a nonlinear system; the solution is mathematically consistent with the equations, which introduce a small fast wave via a \emph{boundary-ponderomotive effect}.

\section{Summary}\label{conc}
In this paper we have addressed the question, \emph{how does the weakly nonlinear Alfv\'en wave behave in the vicinity of a 2D null point?} The null point topology (\ref{eqn:2D_null_point}) and equilibrium variables considered are identical to those considered in the linear study of McLaughlin \& Hood (\citeyear{MH2004}), as is the method for introducing the Alfv\'en wave (i.e. driving the $\hat{\mathbf{z}}$-components of the fluid variables). Thus, we can directly  compare the behaviour of the waves in the linear and nonlinear regimes. 
Our three main results are that
\begin{itemize}
\item[(i)] In $v_z$, the wave propagates along fieldlines at the background Alfv\'en speed, $c_{A}$, accumulating at the separatricies. The wave does not steepen to form a shock (hence, we refer to our choice of $A$ as low amplitude).
\\
\item[(ii)] The Alfv\'en wave sustains cospatial, nonlinear disturbances that have transverse ($v_\perp$, Figure \ref{weakvperp}) and longitudinal ($v_\parallel$, Figure \ref{weakvpar}) manifestations - phenomena not reported before in null point simulations.
\\
\item[(iii)]
During the driving phase, a wave develops in $v_\perp$ and subsequently propagates independently of the Alfv\'en wave. It propagates with the transient properties of a linear fast wave, crossing separatricies and accumulating at the null point.
\end{itemize}

We find that in the low-amplitude limit of the nonlinear solution the majority of the Alfv\'en wave cannot cross separatricies as in the linear solution of McLaughlin \& Hood (\citeyear{MH2004}).
However, we find two key results not seen in the linear case - the wave sustains cospatial daughter disturbances and that an independently propagating fast wave is generated via ponderomotive mode excitation, results not seen in the linear case. 

 The longitudinal daughter, sustained by the Alfv\'en wave, appears to have no real impact upon the medium. The transverse daughter appears to be focused towards $x=0$ separatrix, and thus the null point  as the Alfv\'en wave carries it towards $y=0$. Since the amplitude here is small, such an effect has very little impact on energy transport and dissipation in the vicinity of the null, however the effect has the potential to be significant for larger amplitude Alfv\'en waves.

 A key feature of linear 2D nulls is that the Alfv\'en wave and magnetoacoustic modes are decoupled, and that Alfv\'en wave energy accumulates along the separatricies and not at the null. However, in the nonlinear case we observe the ponderomotive excitation of a fast magnetoacoustic wave which refracts about and eventually accumulates at the null point. Thus, unlike in 2D, some of the Alfv\'en wave's energy is focused at the null point by the fast wave.

After the initial generation of the fast magnetoacoustic wave, we note that no further magnetoacoustic waves are generated, despite the fact that the pulse is travelling through an inhomogeneous region - undergoing longitudinal dispersion ($\nabla_\parallel\, c_A \neq 0$) and phase mixing ($\nabla_\perp \,c_A \neq 0$). The analysis of Thurgood \& McLaughlin (\citeyear{Me2013_PMF}) demonstrated that when a pulse propagates through an inhomogeneous medium, mode conversion can occur, but that it is dependent on the specific scenario (e.g., the phase-mixing experiment of Nakariakov et al. \citeyear{NK97}). As no conversion occurs in our experiment after the initial excitation, $\nabla\,c_A$ in the vicinity of our 2D null point must not be sufficiently steep or sharp enough to further excite magnetoacoustic waves. This suggests that ponderomotive mode conversion due to inhomogeneity will only routinely occur where this profile is sharp or discontinuous. 

\begin{acknowledgements}
The authors acknowledge IDL support provided by STFC. JOT acknowledges travel support provided by the RAS and the IMA, and a Ph.D. scholarship provided by Northumbria University. The computational work for this paper was carried out on the joint STFC and SFC (SRIF) funded cluster at the University of St Andrews (Scotland, UK).
\end{acknowledgements}

\bibliographystyle{aa} 
\bibliography{references.bib}

\end{document}